\documentclass[aps,pra,twocolumn,showpacs,preprintnumbers,amsmath,amssymb,footinbib]{revtex4}
\usepackage{mathtools}

\usepackage{braket}
\usepackage{graphicx,epsfig}
\usepackage{bm}
\usepackage{dcolumn}
\usepackage{xcolor}
\usepackage[breaklinks=true,colorlinks,citecolor=blue,linkcolor=blue,urlcolor=blue]{hyperref}

\def\prv#1#2#3{Phys. Rev. {\bf #1}, #2 (#3)}
\def\rmp#1#2#3{Rev. Mod. Phys. {\bf #1}, #2 (#3)}
\def\prl#1#2#3{Phys. Rev. Lett. {\bf #1}, #2 (#3)}
\def\pra#1#2#3{Phys. Rev. A {\bf #1}, #2 (#3)}

\def\jpamt#1#2#3{J. Phys. A: Math. Theor. {\bf #1}, #2 (#3)}

\def\noi{\noindent}
\def\bc{\begin{center}}
\def\ec{\end{center}}
\newcommand{\bea}{\begin{equation}}
\newcommand{\eea}{\end{equation}\noi}
\newcommand{\ber}{\begin{eqnarray}}
\newcommand{\eer}{\end{eqnarray}\noi}

\begin{document}	

\title{Multi-mode Jaynes-Cummings model results for the collapse and the revival of the quantum Rabi oscillations in a lossy resonant cavity}

\author{Najirul Islam}
\author{Shyamal Biswas}\email{sbsp [at] uohyd.ac.in}

\affiliation{School of Physics, University of Hyderabad, C.R. Rao Road, Gachibowli, Hyderabad-500046, India}

\date{February 7, 2022}

\begin{abstract}
We have numerically obtained theoretical results for the collapse and the revival of the quantum Rabi oscillations for low average number of coherent photons injected on a two-level system in a lossy resonant cavity. We have adopted the multimode Jaynes-Cummings model for the same and especially treated the ``Ohmic" loss to the walls of the cavity, the leakage from the cavity, and the loss due to the spontaneous emission through the open surface of the cavity. We have compared our results with the experimental data obtained by Brune \textit{et al} [\href{http://dx.doi.org/10.1103/PhysRevLett.76.1800}{\prl{76}{1800}{1996}}] in this regard.
\end{abstract}


\maketitle
 

\section{Introduction}
Collapse and revival (CR) of the quantum Rabi oscillations of a two-level system (atom/molecule) is an interesting area of research in the field of cavity quantum electrodynamics \cite{Eberly,Narozhny,Knight,Puri,Rempe,Haroche-1998,Brune,Raimond,Haroche,Berman,Chong}. Eberly \textit{et al} first predicted the phenomenon of the CR within the single-mode Jaynes-Cummings (J-C) model \cite{Jaynes} for the quantum Rabi oscillations of a two-level system interacting with coherent photons in a cavity \cite{Eberly}. The CR was subsequently observed by investigating the dynamics of the interaction of a single Rydberg atom with the resonant mode of an electromagnetic field in a superconducting cavity \cite{Rempe}. The CR may find applications in supersymmetric qubits \cite{Chong}. While the existing theories \cite{Eberly,Narozhny,Knight,Puri,Berman,Chong} for the CR usually require a large average number of photons ($\bar{n}\gg1$) in the coherent field, a seminal experiment \cite{Brune} on the same was carried out by Brune \textit{et al} for a low average number of photons ($\bar{n}\gtrsim1$) in the coherent field. In fact, as far as we know, all the experiments on the CR were carried out for low average number of photons \cite{Rempe,Meekhof,Brune} except the one \cite{Meunier} carried out for $\bar{n}=13.4$. Hence we theoretically investigate the CR for a low average number of photons in a coherent field. Theory for the CR is also available for low average number of injected coherent photons as well as for all values of the average number of the injected coherent photons \cite{Yoo,Haroche-1998,Meystre,Brune,Meunier,Azuma,Berman}. This theory takes only the resonant mode into account for the light-matter interactions. We are, however, interested in considering multi-modes into account. 

\begin{figure}
\includegraphics[width=1.00 \linewidth]{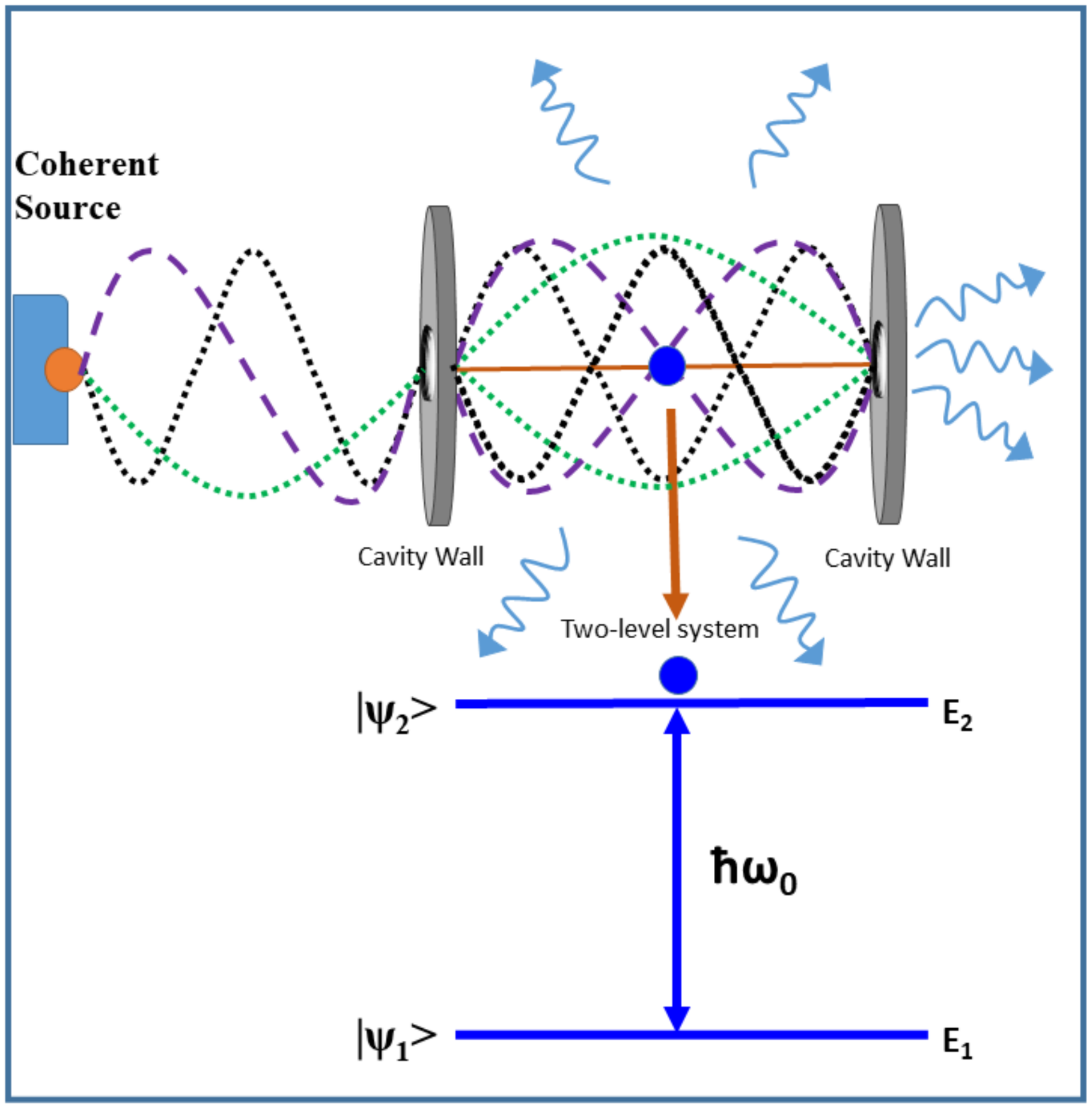}
\caption{Schematic diagram for a two-level system interacting with injected coherent photons in a lossy resonant cavity.
\label{fig2}}
\end{figure}

J-C model takes only the resonant cavity mode into account for the explanation of the CR of the quantum Rabi oscillations of a two-level system in a loss-less cavity \cite{Jaynes,Eberly}. However, the cavities are not loss-less in reality \cite{Brune}. This brings a frequency broadening as well as the appearance of multi-modes around the resonant mode into account. Brune \textit{et al}'s experiment on the CR were carried out in a lossy resonant cavity of the mode quality factor $Q=7\times10^7$  \cite{Brune}. The schematic diagram for the two-level system interacting with the injected coherent photons in the lossy resonant cavity is shown in figure \ref{fig2}. It is clear from figure \ref{fig2} how the injected coherent photons are introduced into the cavity and how the two-level system is interacting with the multi-modes of the injected coherent photons in the cavity. Losses from the cavity are shown by the wavy arrows in the same figure. The frequency broadening in Brune \textit{et al}'s experiment can be attributed to the multi-mode J-C model, $\hat{H}=\frac{1}{2}\hbar\omega_0\sigma_3+\sum_{\vec{k}s}\hbar\omega_{\vec{k}}\hat{a}_{\vec{k}s}^\dagger\hat{a}_{\vec{k}s}-i\sum_{\vec{k}s}\hbar g_{\vec{k}s}[\sigma_+\hat{a}_{\vec{k}s}-\sigma_-\hat{a}_{\vec{k}s}^\dagger]$\footnote{Here $\hat{H}$ is the Hamiltonian operator for the two-level system interacting with the injected coherent photons in a lossy resonant cavity.  We are following the notation \cite{Lahiri,Islam}: $\sigma_+=\ket{\psi_2}\bra{\psi_1}$, $\sigma_-=\ket{\psi_1}\bra{\psi_2}$, $\sigma_1=[\sigma_++\sigma_-]$, $\sigma_2=-i[\sigma_+-\sigma_-]$ and $\sigma_3=\ket{\psi_2}\bra{\psi_2}-\ket{\psi_1}\bra{\psi_1}$ where $\ket{\psi_1}$ ($\ket{\psi_2}$) is the energy eigenstate for the lower (higher) energy $E_1$ ($E_2$) of the two-level system for no light-matter interactions and $\omega_0=(E_2-E_1)/\hbar$ is the Bohr frequency of the two-level system. We further have $\hat{a}_{\vec{k}s}$ ($\hat{a}_{\vec{k}s}^\dagger$) which annihilates (creates) a photon of energy $\hbar\omega_{\vec{k}}$, polarization $s$ and momentum $\hbar\vec{k}$ in the Fock space. We also have $g_{\vec{k}s}$ as the real coupling constant for the light-matter interaction for the mode $\vec{k}s$.} \cite{Seke,Lahiri}, rather than the single-mode Jaynes-Cummings model \cite{Islam}. Thus the theoretical explanation of the CR of the quantum Rabi oscillations in a lossy resonant cavity needs a novel approach with the multi-mode Jaynes-Cummings model. The novel approach must take losses from the cavity into account for the explanation of Brune \textit{et al}'s experimental data \cite{Brune}. Here we provide a novel theory for the CR by considering losses from the cavity especially the ``Ohmic" loss \cite{Siegman} to the walls of the cavity, the leakage from the cavity, and the loss due to the spontaneous emission through the open surface of the cavity.

Multi-mode J-C model \cite{Seke} is well-known \cite{Islam,Li,Shen} as an extension of the single-mode J-C model \cite{Jaynes}. Multi-mode J-C model has been successfully used by us \cite{Islam} to explore the quantum Rabi oscillations of a two-level system interacting with a very low average number of injected coherent photons ($\bar{n}=0.4$) in a lossy resonant cavity as described in Brune \textit{et al}'s experiment \cite{Brune}. Such a very low average number of photons was treated perturbatively (up to the second order in $\bar{n}^2\ll1$) in Ref. \cite{Islam}. However, Brune \textit{et al} \cite{Brune} obtained two more sets of data for low average number of injected coherent photon numbers  $\bar{n}=0.85\pm0.04$ and $\bar{n}=1.77\pm0.15$ in the same cavity showing the CR of the quantum Rabi oscillations of a two-level system ($^{87}$Rb atom). A non-perturbative method is needed for the theoretical explanation of these two sets of data for the CR. Hence we extend our method described in Ref. \cite{Islam} for this purpose. The CR of the quantum Rabi oscillations, of course, were not discussed in Ref. \cite{Islam}.

Calculation in this article essentially begins with Eqn. (8) of Ref. \cite{Islam}. This equation is an outcome of the multi-mode J-C model and it is nothing but the net transition probability ($P_{2\rightarrow1}(t)$) which describes the quantum Rabi oscillations in time ($t$) domain for a two-level system interacting with coherent photons in a lossy resonant cavity. This transition probability is a function of time and a number of parameters including the renormalized coupling constant which can be determined by the mode quality factor of the cavity and the average number of coherent photons incident on the two-level system. We determine the transition probabilities for the average numbers of coherent photons $\bar{n}=0.85$ and $\bar{n}=1.77$ and the mode quality factor $Q=7\times10^7$ after determining the renormalized coupling constants within a graphical method. We compare our theoretical results with the experimental data obtained by Brune \textit{et al} \cite{Brune} and the existing theoretical results obtained within the single-mode J-C model \cite{Brune,Raimond,Meunier}. We also estimate the collapse time and the revival time for $\bar{n}=0.85$ and $1.77$. Finally, we conclude.

\section{Collapse and Revival}
Let us consider a two-level system (atom/molecule) in a lossy resonant optical cavity of the resonance frequency $\omega_0$ and the mode quality factor $Q$. The two-level system is interacting with the coherent photons which are injected through a hole on the cavity axis. Let the average number of coherent photons injected on the two-level system be $\bar{n}$. We consider the quantum Rabi oscillations of the two-level system in the processes of the spontaneous emission, the stimulated emission and the stimulated absorption. The quantum Rabi oscillations need the two-level system to strongly interact with the injected photons of the cavity field. The high mode quality factor of the cavity ensures strong light-matter coupling. The photon emitted from the two-level has a long life-time ($\sim200~\mu$s) in such a situation. The emitted photon repeatedly reflects back and forth with the mirrors of the cavity before it leaks out through the holes on the axis of the cavity or becomes absorbed (or scattered) in the walls of the cavity resulting in the ``Ohmic" loss \cite{Siegman}. However, the curved surface of the cylindrical geometry of the cavity is also kept open. This causes additional loss from the cavity. This loss is associated with the spontaneous emission from the two-level system through the curved surface of the cavity \cite{Islam}. The probability that an emitted photon escapes from the cavity through the curved surface is $p_0=\frac{2\pi rh}{2\pi rh+2\pi r^2}=\frac{1}{1+\frac{r}{h}}$ where $r$ is the radius of each of the mirrors of the cavity and $h$ is the separation of the two mirrors \cite{Islam}. All these losses result in the net quality factor as $Q'=\frac{1}{\frac{1}{Q}+\frac{p_0A(0)}{\omega_0}}$ \cite{Islam} where $A(0)$ is the frequency broadening due to the natural decay in the free space inside the cavity and $\omega_0$ is the Bohr frequency of the two-level system. Here $A(0)$ is nothing but the enhanced value of the Einstein $A$ coefficient due to the Purcell effect \cite{Purcell}. Derivation of the net quality factor has been shown in Appendix \ref{A}. Let us consider that initially ($t=0$) the two-level system was in the excited state. Thus we get the net transition probability of two-level system from the excited state ($\ket{\psi_2}$) to the ground state ($\ket{\psi_1}$) at time $t$, as \cite{Islam}
\begin{eqnarray}\label{eqn:1}
P_{2\rightarrow1}(t)&=&A(0)\sum_{n=0}^\infty \frac{4}{\pi}\frac{\bar{n}^n}{n!}\text{e}^{-\bar{n}}[n+1]\times\nonumber\\&&\int_{\omega_{n}}^\infty\frac{(\omega_0/Q')^2}{4[\Omega_{n}^2-\omega_{n}^2]+({\omega_0}/{Q'})^2}\frac{\sin^2(\Omega_{n}t/2)}{\Omega_{n}\sqrt{\Omega_{n}^2-\omega_{n}^2}}\text{d}\Omega_{n}\nonumber\\
\end{eqnarray}
where $\omega_{n}=2[n+1]^{1/2}g_{\omega_0}'(\bar{n},Q')$ is the renormalized $n$-photon Rabi frequency, $\frac{\bar{n}^n}{n!}\text{e}^{-\bar{n}}$ represents the probability ($p_n$) of occupation of $n$ coherent photons, and $g_{\omega_0}'(\bar{n},Q')$ is the renormalized light-matter coupling constant. Eqn.~(\ref{eqn:1}) is a special case of Eqn. (8) of Ref. \cite{Islam}. Re-derivation of Eqn.~(\ref{eqn:1}) has been outlined in Appendix \ref{B}.   

\begin{figure}
\includegraphics[width=1.00 \linewidth]{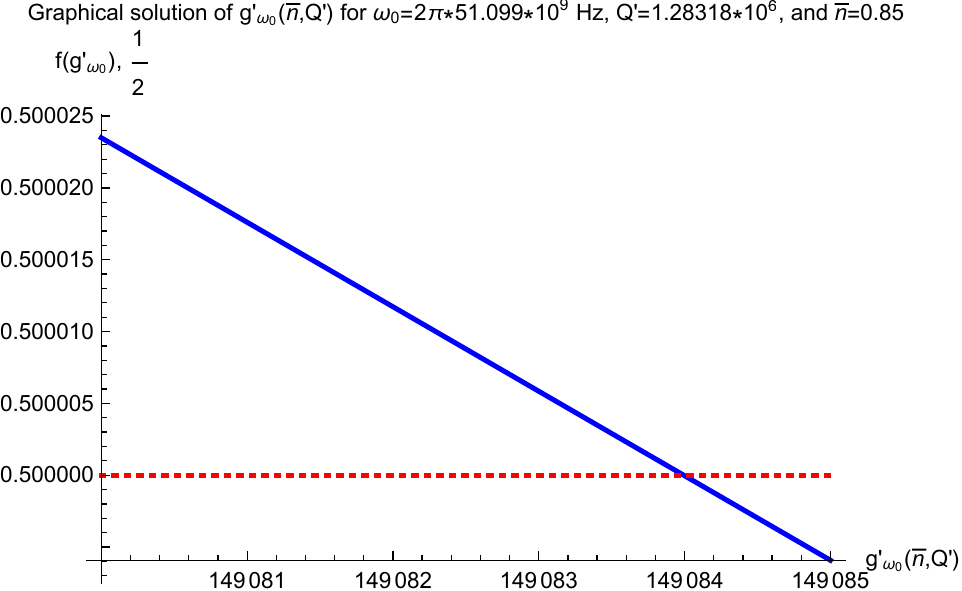}
\caption{Solid line represents the left-hand side of Eqn. (\ref{eqn:2}) for the parameters as mentioned in the plot label and for $A(0)=0.473053\times10^6$~Hz. Dotted line represents the right-hand side of the same equation for the same parameters.
\label{fig0}}
\end{figure}

Eqn. (\ref{eqn:1}), however, is able to describe the CR of the quantum Rabi oscillations of the two-level system in the lossy resonant cavity for the low average number of coherent photons ($\bar{n}\gtrsim1$). The descriptions of the CR require determination of the renormalized coupling constant which, however, was not done in any literature for the low average number of coherent photons ($\bar{n}\gtrsim1$). Let us now describe the CR in terms of the multi-mode J-C model.

For $t\rightarrow\infty$, we have $\sin^2(\Omega_{n}t/2)\rightarrow\frac{1}{2}$ in Eqn. (\ref{eqn:1}). Experimental observation suggests us to take the limiting value $\lim_{t\rightarrow\infty}P_{2\rightarrow1}(t)=1/2$ \cite{Brune,Meekhof}. Now by setting $P_{2\rightarrow1}(\infty)=1/2$ and integrating over $\Omega_{n}$ in Eqn. (\ref{eqn:1}), we get
\begin{eqnarray}\label{eqn:2}
f(g'_{\omega_0})=\frac{1}{2}
\end{eqnarray}
where
\begin{eqnarray}\label{eqn:2a}
f(g'_{\omega_0})=\sum_{n=0}^{\infty}\frac{A(0)[\omega_0/Q']\text{e}^{-\bar{n}}\bar{n}^n}{2g'_{\omega_0}(\bar{n},Q')[4g'_{\omega_0}(\bar{n},Q')+\frac{\omega_0}{Q'\sqrt{1+n}}]n!}.
\end{eqnarray}
The Einstein $A$ coefficient was perturbatively calculated from the experimental data \cite{Brune} for the `vacuum' Rabi oscillation of the two-level system ($^{87}$Rb atom) for the Rabi frequency $\Omega_R=2\pi\times47\times10^3$Hz, the Bohr frequency as well as resonance frequency $\omega_0=2\pi\times51.099\times10^9$ Hz, the average number of thermal photons $\bar{n}=0.0489$, and the cavity specification $r=25~$mm, $h=27~$mm and $Q=7\times10^7$, as $A(0)=0.473053\times10^6~$Hz \cite{Islam}. The net quality factor is resulted in as $Q'=1.28318\times10^6$ for of all these parameters \cite{Islam}. The experiment on the CR of the quantum Rabi oscillations of the same two-level system was done in the same setup except the thermal photons replaced with the injected coherent photons. Such a replacement does not change the Einstein $A$ coefficient rather changes the renormalized coupling constant $g'_{\omega_0}(\bar{n},Q')$ as well as the Rabi frequency $\Omega_R=2[\bar{n}+1]^{1/2}g'_{\omega_0}(\bar{n},Q')$ \cite{Islam}. The perturbation method, as employed in Ref. \cite{Islam}, though suits for very low average number of thermal or coherent photons, does not suit for average number of photons greater than or comparable to $1$.

\begin{figure}
\includegraphics[width=0.98 \linewidth]{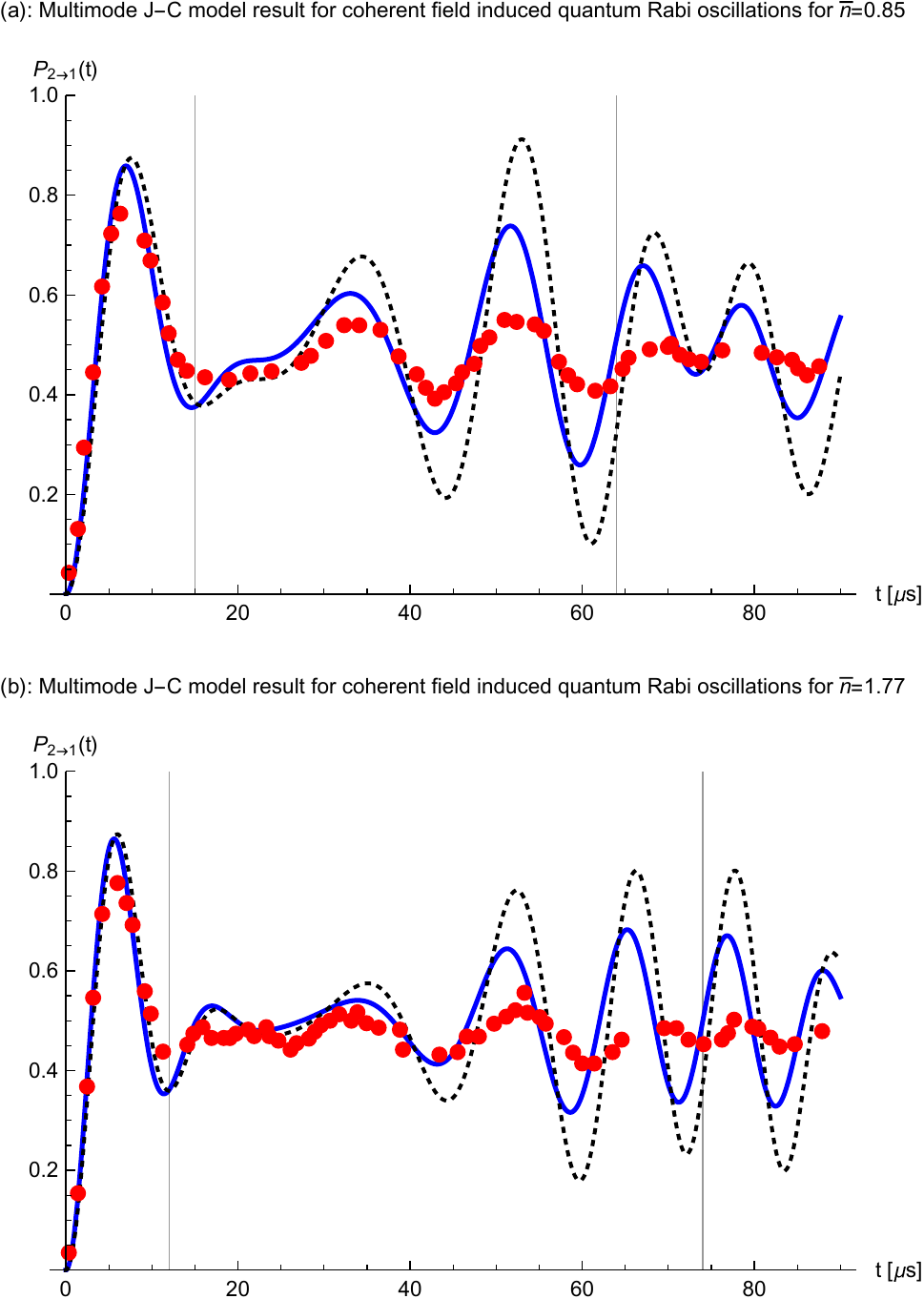}
\caption{
(a): Solid line follows Eqn.~(\ref{eqn:1}) for $\omega_0=2\pi\times51.099\times10^9~$Hz, $g'_{\omega_0}=149084$~Hz, $\Omega_R=2\pi\times64.5457\times10^3~$Hz, $\bar{n}=0.85$, and $Q'=1.28318\times10^6$. Circles represent adapted experimental data obtained by Brune \textit{et al} \cite{Brune} in this regard for the circular Rydberg states (with the principal quantum numbers $50$ and $51$) of the $^{87}$Rb atoms. Dotted line follows Eqn.~(\ref{eqn:3}) for $g=g'_{\omega_0}=149084$~Hz and $\bar{n}=0.85$.\\
(b): Solid line follows Eqn.~(\ref{eqn:1}) for $\omega_0=2\pi\times51.099\times10^9~$Hz, $g'_{\omega_0}=152852$~Hz, $\Omega_R=2\pi\times80.9769\times10^3~$Hz, $\bar{n}=1.77$, and $Q'=1.28318\times10^6$. Circles represent adapted experimental data obtained by Brune \textit{et al} \cite{Brune} in this regard for the circular Rydberg states (with the principal quantum numbers $50$ and $51$) of the $^{87}$Rb atoms. Dotted line follows Eqn.~(\ref{eqn:3}) for $g=g'_{\omega_0}=152852$~Hz and $\bar{n}=1.77$.
\label{fig1}}
\end{figure}

Let us now determine the renormalized coupling constant $g'_{\omega_0}(\bar{n},Q')$ from Eqn. (\ref{eqn:2}) for fixed $\bar{n}$ and $A(0)=0.473053\times10^6~$Hz, $\omega_0=2\pi\times51.099\times10^9$ Hz and $Q'=1.28318\times10^6$ as mentioned above. We already have mentioned that Brune \textit{et al} took $\bar{n}=0.85\pm0.04$ and $1.77\pm0.15$ for the observations of the CR \cite{Brune}. We employ the graphical method for the determination of the renormalized coupling constant from  Eqn. (\ref{eqn:2}) for the above parameters. This method is considered to be a non-perturbative method. We plot both the left-hand side (solid line) and the right-hand side (dotted line) of Eqn. (\ref{eqn:2}) with respect to the renormalized coupling constant in figure \ref{fig0} for the above parameters and $\bar{n}=0.85$. The intersection of these two plots solves the renormalized coupling constant on the horizontal axis of figure \ref{fig0}. Thus for $\bar{n}=0.85$ we get the renormalized coupling constant to be as $g'_{\omega_0}=149084$~Hz. This graphical solution can be called as a numerical solution because the left-hand side of Eqn. (\ref{eqn:2}) is evaluated numerically before being plotted in figure \ref{fig0}. Similarly, for the same parameters except for $\bar{n}=1.77$ we get the renormalized coupling constant to be as $g'_{\omega_0}=152852$~Hz. Value of the renormalized coupling constant enables us to plot Eqn. (\ref{eqn:1}) where periodic dephasing over the time for various photon numbers ($n$) results in the collapse in the quantum Rabi oscillations and periodic rephasing over the time for various photon numbers results in revival in the quantum Rabi oscillations. Multi-modes further result in dephasing in the quantum Rabi oscillations. We plot Eqn. (\ref{eqn:1}) in figure \ref{fig1}-a for $\bar{n}=0.85$ and in figure \ref{fig1}-b for $\bar{n}=1.77$. We compare our results (solid lines) with the corresponding sets of experimental data in figures \ref{fig1}-a and \ref{fig1}-b. We also need to compare our results with the existing theoretical result \cite{Brune,Meunier,Raimond}
\begin{eqnarray}\label{eqn:3}
P_{2\rightarrow1}(t)=\sum_{n=0}^{\infty}\frac{\text{e}^{-\bar{n}}\bar{n}^n}{n!}\sin^2\big(gt\sqrt{n+1}\big)
\end{eqnarray}
obtained for similar purpose under the consideration of the light-matter coupling with only the resonant mode ($\omega=\omega_0$) and no loss from the cavity. Dotted lines in figure \ref{fig1}-a and \ref{fig1}-b follow Eqn.~(\ref{eqn:3}) and represent single-mode J-C model results for the coherent field-induced quantum Rabi oscillations for the same coupling constants $\{g=g'_{\omega_0}\}$ and the same average photon numbers $\{\bar{n}\}$ taken for the solid lines.

The collapse happens at a point ($t=t_c$) when different quantum Rabi oscillations take place in $\pi$ amount of out of phase. This causes destructive interference in the quantum Rabi oscillations. For low average number of coherent photons ($\bar{n}\gtrsim1$) too, the maximum of the Poisson distribution $p_n=\frac{\bar{n}^n}{n!}\text{e}^{-\bar{n}}$ in Eqn.~(\ref{eqn:1}) occurs at around $n=\bar{n}$. The standard deviation for the distribution is $\triangle n=\sqrt{\bar{n}}$. Thus we apply the condition ($[\omega_{\bar{n}+\triangle n}-\omega_{\bar{n}-\triangle n}]t_c=\pi$ \cite{Ficek}) for the collapse in Eqn.~(\ref{eqn:1}), as  
\begin{eqnarray}\label{eqn:4}
2g'_{\omega_0}(\bar{n},Q')\bigg[\sqrt{\bar{n}+\sqrt{\bar{n}}+1}-\sqrt{\bar{n}-\sqrt{\bar{n}}+1}\bigg]t_c=\pi.
\end{eqnarray}
Here-from we estimate the collapse times as $t_c\simeq15~\mu$s for the first set of data (for figure \ref{fig1}-a) and $t_c\simeq12~\mu$s for the second set of data (for figure \ref{fig1}-b). Vertical lines at $t=15~\mu$s and $12~\mu$s indicate the collapse of the quantum Rabi oscillations in figures \ref{fig1}-a and \ref{fig1}-b, respectively.

The revival takes place at a point ($t=t_r$) when all the neighbouring quantum Rabi oscillations come in phase again and add up for constructive interference. Thus we apply the condition ($[\omega_{\bar{n}+1}-\omega_{\bar{n}}]t_r=2\pi$ \cite{Meunier,Berman,Ficek}) for the revival in Eqn.~(\ref{eqn:1}), as  
\begin{eqnarray}\label{eqn:5}
2g'_{\omega_0}(\bar{n},Q')\bigg[\sqrt{\bar{n}+2}-\sqrt{\bar{n}+1}\bigg]t_r=2\pi.
\end{eqnarray}
Here-from we estimate the revival times as $t_r\simeq64~\mu$s for the first set of data (for figure \ref{fig1}-a) and $t_r\simeq74~\mu$s for the second set of data (for figure \ref{fig1}-b). Vertical lines at $t=64~\mu$s and $74~\mu$s indicate the revival of the quantum Rabi oscillations in figures \ref{fig1}-a and \ref{fig1}-b, respectively.

\section{Conclusion}
We have theoretically obtained multi-mode Jaynes-Cummings model results for the CR of the quantum Rabi oscillations of a two-level system interacting with injected coherent photons in a lossy resonant cavity. We have extended our previous theory \cite{Islam} within a non-perturbative method in this regard. We have compared our results with two sets of experimental data \cite{Brune} for low average number of coherent photons ($\bar{n}=0.85$ and $1.77$) incident on a two-level system in the lossy resonant cavity. Our results match reasonably well with the experimental data, at least, better than the existing theoretical one (Eqn.~(\ref{eqn:3})) \cite{Brune,Meunier,Raimond} obtained for only the resonant mode and no loss from the cavity under consideration.

We had to cite Ref. \cite{Islam} quite often in this article because it is an extension of the previous work \cite{Islam} on the quantum Rabi oscillations. This extension is necessary because the CR of the quantum Rabi oscillations of a two-level system, however, has separate existence \cite{Eberly,Narozhny,Knight,Puri,Rempe,Berman,Chong,Meunier,Yoo,Azuma} over the usual discussions on the quantum Rabi oscillations.

The solid line, which represents the function $f(g'_{\omega_0})$ of the renormalized coupling constant in the plot of figure \ref{fig0}, appears to be straight in the plot for the small range of the renormalized coupling constant. It would not have appeared to be a straight line if we had taken a large range of the renormalized coupling constant in the plot.

It is clear from figure \ref{fig1} that the multi-mode Jaynes-Cummings model result is almost the same as the single-mode Jaynes-Cummings model result for short time-evolution of the net transition probability. These two results significantly differ at a large time. This implies that the non-resonant modes are significant at large times.

The net transition probability ($P_{2\rightarrow1}(t)$ in Eqn. (\ref{eqn:1})) represents a dynamical behaviour of the two-level system. The dynamical behaviour of the system can be analysed in the Schr$\ddot{\text{o}}$dinger picture, the Heisenberg picture, and the interaction picture. However, the multi-mode J-C Hamiltonian mentioned in the introductory section has been expressed in the Schr$\ddot{\text{o}}$dinger picture. This picture has an advantage of making the Hamiltonian operator to be time-independent and evolving the quantum mechanical state with respect to time for the given Hamiltonian.

We have developed a theory for the CR with an emphasis for the low average number of injected coherent photons ($\bar{n}$) in a lossy resonant cavity. Eqns.~(\ref{eqn:1}) and (\ref{eqn:2}) are our key results in this regard. However, nowhere in these two equations, even in the subsequent equations, we have considered $\bar{n}$ to be small. Hence our theory is applicable for all values of the average number of injected coherent photons.

Our results are significantly different from the previous theoretical results \cite{Brune,Meunier,Raimond} from (i) the consideration of the multi-modes around the resonant mode into account, and (ii) the consideration of the frequency broadening due to the ``Ohmic" loss to the walls of the cavity, the leakage from the cavity, and the loss due to the spontaneous emission through the open surface of the cavity. However, our theory did not match well with the experimental data in the region where the time $t$ is large, say, for $t\gtrsim40~\mu$s in figures \ref{fig1}-a and \ref{fig1}-b. It is clear from these two figures that the amplitude of the oscillation of the net transition probability ($P_{2\rightarrow1}(t)$) need to be smaller than that we have predicted for large $t$. From the Fourier transform of the Lorentzian distribution, we know that the Lorentzian broadening of the frequencies of harmonic oscillations causes an exponential decay of the amplitude of the net oscillation in the time domain. Thus the amplitude of the oscillation of the net transition probability would be smaller at a larger time if additional losses, which cause additional frequency broadening, are taken into account. Further consideration of substantial losses corresponding to the frequency broadening due to the inhomogeneous light-matter coupling, Doppler broadening, thermal broadening, \textit{etc} may improve our theory. Such an improvement is kept as an open problem.

\section*{Acknowledgement}
S. Biswas acknowledges partial financial support of the SERB, DST, Govt. of India under the EMEQ Scheme [No. EEQ/2019/000017]. We acknowledge useful discussions with Dr. V. Ashoka (UoH, Hyderabad) and Dr. P. Prem Kiran (UoH, Hyderabad).  We thank Mr. Pawan Kumar Verma (UoH, Hyderabad) for helping us in drawing figure \ref{fig2}. We also thank the anonymous reviewers for their thorough review. We highly appreciate their comments which significantly contributed to improving the quality of the paper.

\appendix
\section{Derivation of the net quality factor}
\label{A}
The quality factor $Q$ of a resonant optical cavity is defined as follows \cite{Christopoulos}
\begin{eqnarray}\label{eqn:A1}
Q=w_0\frac{W}{P_{loss}}
\end{eqnarray}
where $w_0$ is the (angular) resonance frequency, $W$ is the electromagnetic energy stored in the resonant mode of the cavity, and $P_{loss}$ is the electromagnetic energy lost per optical cycle to the walls of the cavity. 

The ``Ohmic" loss, apart from the loss due to the surface a.c. current flow in the (conducting) cavity walls, also includes the losses due to the host-crystal absorption, impurities, scattering loss, excited-state absorption, and other effects \cite{Siegman}. We are also considering the loss of the stored electromagnetic energy due to the leakage from the holes on the cavity axis in addition to the ``Ohmic" loss. If $Q_1$ be the quality factor of the cavity corresponding to the ``Ohmic" loss and $Q_2$ be the quality factor of the cavity corresponding to the leakage, then the inverse of quality factor ($Q$) of the cavity corresponding to both the losses is given by \cite{Christopoulos}
\begin{eqnarray}\label{eqn:A2}
\frac{1}{Q}=\frac{1}{Q_1}+\frac{1}{Q_2}.
\end{eqnarray}
The ``Ohmic" loss, however, would be very low for a superconducting cavity and the quality factor for such a cavity would be very high ($Q=7\times10^7$ \cite{Brune}). The differential equation for the variation of the stored electromagnetic energy over the time $t$ follows \cite{Christopoulos,Thyagarajan}
\begin{eqnarray}\label{eqn:A3}
\frac{\text{d}W}{\text{d}t}=-\frac{\omega_0}{Q}W
\end{eqnarray}
for both the ``Ohmic" loss and the leakage together.
A solution to this equation is
\begin{eqnarray}\label{eqn:A4}
W(t)=W(0)\text{e}^{-\frac{\omega_0t}{Q}}.
\end{eqnarray}
Here-from we can write the temporal part of the electric field associated with the resonant mode, as \cite{Thyagarajan}
\begin{eqnarray}\label{eqn:A5}
E(t)=E_0\text{e}^{-\frac{\omega_0t}{2Q}}\text{e}^{-i\omega_0t}.
\end{eqnarray}
It is clear from the above equation that the oscillation of the electric field dies as $E(t)\propto E_0\text{e}^{-\frac{\omega_0t}{2Q}}$ in the resonant optical cavity. Fourier transform of the above temporal part of the electric field becomes
\begin{eqnarray}\label{eqn:A6}
\tilde{E}(\omega)=E_0\int_0^\infty\text{e}^{-\frac{\omega_0t}{2Q}}\text{e}^{i[\omega-\omega_0]t}\text{d}t=\frac{E_02iQ}{2Q[\omega-\omega_0]+i\omega_0}~~~
\end{eqnarray}
for the time $t>0$ and the (angular) frequency $\omega>0$. Here-from we get the spectral distribution of the (angular) frequencies, as \cite{Thyagarajan}
\begin{eqnarray}\label{eqn:A7}
|\tilde{E}(\omega)|^2=E_0^2\frac{1}{[\omega-\omega_0]^2+\frac{\omega_0^2}{4Q^2}}.
\end{eqnarray}
The shape of this spectral distribution is Lorentzian and the full width at half maximum (FWHM) of the distribution is given by $\triangle\omega=\frac{\omega_0}{Q}$ \cite{Thyagarajan}.

Another Lorentzian broadening of the (angular) frequencies, similar to the one in Eqn. (\ref{eqn:A7}), is also obtained for the spontaneous emission from a two-level system (atom/molecule) in the free space within the Weisskopf-Wigner approximation, as \cite{Weisskopf,Thyagarajan2}
\begin{eqnarray}\label{eqn:A8}
|\tilde{E}(\omega)|^2=E_0^2\frac{1}{[\omega-\omega_0]^2+\frac{A(0)^2}{4}}
\end{eqnarray}
where $A(0)$ is the Einstein $A$ coefficient. The width (FWHM) of this Lorentzian broadening is given by $\triangle\omega=A(0)$ \cite{Thyagarajan}. However, if the two-level system be kept in the cavity, then only a fraction of the total spontaneous emission can escape from the cavity resulting in an additional loss through the open surface of the cavity. Let the cavity be of cylindrical shape and its curved surface is open. The probability that an emitted photon escapes from the cavity through the curved surface is $p_0=\frac{2\pi rh}{2\pi rh+2\pi r^2}=\frac{1}{1+\frac{r}{h}}$ where $r$ is the radius of each of the mirrors of the cavity and $h$ is the separation of the two mirrors \cite{Islam}. Thus Eqn. (\ref{eqn:A8}) would be modified for the spontaneous emission through the curved surface of the cavity, as
\begin{eqnarray}\label{eqn:A9}
|\tilde{E}(\omega)|^2=E_0^2\frac{1}{[\omega-\omega_0]^2+\frac{p_0^2A(0)^2}{4}}.
\end{eqnarray}
Convolution of the two Lorentzian distributions of Eqns. (\ref{eqn:A7}) and (\ref{eqn:A9}) is also another Lorentzian distribution with the net width (FWHM) $\triangle\omega'=\frac{\omega_0}{Q}+p_0A(0)$ which is the addition of the widths (FWHM) of the two distributions \cite{Fultz}. If we compare Eqn. (\ref{eqn:A9}) with Eqn. (\ref{eqn:A7}) then we can assign a quality factor for the loss associated with the spontaneous emission through the curved surface of the cavity as
\begin{eqnarray}\label{eqn:A10}
Q_3=\frac{\omega_0}{p_0A(0)}. 
\end{eqnarray}
 The inverse of the net quality factor corresponding to the broadenings of Eqns. (\ref{eqn:A7}) and (\ref{eqn:A9}), on the other hand, would be an addition of the inverse of the individual quality factors as mentioned in Eqn. (\ref{eqn:A2}) \cite{Christopoulos}. Thus the inverse of the net quality factor of the lossy resonant optical cavity of our interest would be $\frac{1}{Q'}=\frac{1}{Q}+\frac{1}{Q_3}=\frac{1}{Q}+\frac{p_0A(0)}{\omega_0}$. Here-from we get the desired net quality factor, as \cite{Islam}
\begin{eqnarray}\label{eqn:A11}
 Q'=\frac{1}{\frac{1}{Q}+\frac{p_0A(0)}{\omega_0}}.
\end{eqnarray}
This form of the net quality factor has been used in Eqn. (\ref{eqn:1}).

\section{Re-derivation of Eqn. (\ref{eqn:1})}
\label{B}
The J-C model result for the probability of stimulated or spontaneous emission of a photon from the two-level system which is initially ($t=0$) found in the excited state ($\ket{\psi_2}$) in the cavity, takes the form within the electric dipole approximation at time $t$, as \cite{Jaynes,Lahiri,Islam}
\begin{eqnarray}\label{eqn:1a}
P_{2\rightarrow1}^{n\rightarrow n+1}(g_{\vec{k}s}, \omega_{\vec{k}}, t)&=&4g_{\vec{k}s}^2\times[n+1]\nonumber\\&&\times\frac{\sin^2\big(\frac{\sqrt{[\omega_{\vec{k}}-\omega_0]^2+4g_{\vec{k}s}^2[n+1]} t}{2}\big)}{[\omega_{\vec{k}}-\omega_0]^2+4g_{\vec{k}s}^2(n+1)}.~~~~~~
\end{eqnarray}
Here $\omega_{\vec{k}}$, $\vec{k}$ and $s$ are the angular frequency, wavevector and polarization of the emitted photon, respectively, over $n$ such identical photons in the cavity, $\omega_0$ is Bohr frequency of the system having the ground state $\ket{\psi_1}$, $g_{\vec{k}s}=\langle\psi_1|\hat{\vec{d}}\cdot\hat{\text{e}}_{\vec{k}s}|\psi_2\rangle\sqrt{\frac{\omega_{\vec{k}}}{2\hbar\epsilon_0V}}$ \cite{Seke} is the coupling constant for the light-matter interaction, $\hat{\vec{d}}$ is the electric dipole moment operator for the system, $\hat{\text{e}}_{\vec{k}s}$ is the unit-vector for the polarization of the cavity field, and $V$ is the effective volume of the cavity \cite{Jaynes,Lahiri,Islam}. Eqn. (\ref{eqn:1a}) describes the quantum Rabi oscillations of the two-level system in the cavity. Incident light, however, makes back and forth reflections with the walls of the cavity. Polarization of the cavity field does not change under such reflections. The wavevector, however, changes the sign under the reflection. The above transition probability (i.e. Eqn. (\ref{eqn:1a})) can now be expressed in the frequency ($\omega$) domain, as
\begin{eqnarray}\label{eqn:1b}
P_{2\rightarrow1}^{n\rightarrow n+1}(g_{\omega}, \omega, t)&=&4g_{\omega}^2\times[n+1]\nonumber\\&&\times\frac{\sin^2\big(\frac{\sqrt{[\omega-\omega_0]^2+4g_{\omega}^2[n+1]} t}{2}\big)}{[\omega-\omega_0]^2+4g_{\omega}^2[n+1]}~~~
\end{eqnarray}
where $\omega_{\vec{k}}$ is replaced with $\omega$ and $g_{\vec{k}s}$ is replaced with the new coupling constant $g_{\omega}$ (such that $g_{\omega}^2=\langle g_{\vec{k}s}^2\rangle$) once the averaging over the two opposite directions of the wavevector $\vec{k}$ and $-\vec{k}$ is done.

Let us now consider a coherent electromagnetic field be injected on the two-level system in the cavity. Frequency broadening of the coherent field results in energy loss from the cavity. For multi-modes as well as for all possible frequencies of the injected coherent field, the net transition probability would be an integration of the right-hand side of Eqn. (\ref{eqn:1b}) over the frequency ($\omega$) and summation over the number of photons with the proper weightage ($p_n(\omega)$) of the occupation probability of the coherent photons, as \cite{Islam}   
\begin{eqnarray}\label{eqn:1c}
P_{2\rightarrow1}(t)&=&A(0)\sum_{n=0}^\infty [n+1]\int_0^\infty p_n(\omega)\nonumber\\&&\times\frac{\sin^2\big(\frac{\sqrt{[\omega-\omega_0]^2+4g_{\omega}'^2(n+1)} t}{2}\big)}{[\omega-\omega_0]^2+4g_{\omega}'^2[n+1]}\text{d}\omega
\end{eqnarray}
where $A(0)$ is the Einstein $A$ coefficient and $g_{\omega}'$ is the renormalized coupling constant which replaces $g_{\omega}$ and takes care of the limit $P_{2\rightarrow1}(\infty)=1/2$ \cite{Brune}. The Einstein coefficient has appeared from the definition that $\lim_{t\rightarrow0}|\frac{d}{dt}P_{2\rightarrow1}(t)|=A(0)$ for no incident photons \cite{Islam}. Eqn. (\ref{eqn:1c}) is a multi-mode J-C model result. Since most of the contributions of the integration in Eqn. (\ref{eqn:1c}) are coming from around the resonance ($\omega\rightarrow\omega_0$), we replace $g_{\omega}'$ with $g_{\omega_0}'$ and $p_n(\omega)$ with $p_n(\omega_0)$ to recast Eqn. (\ref{eqn:1c}) within the rotating wave approximation ($\omega+\omega_0\gg|\omega-\omega_0|$ or $\omega_0^2\gg4g_{\omega_0}'^2$), as
\begin{eqnarray}\label{eqn:1d}
P_{2\rightarrow1}(t)=2A(0)\sum_{n=0}^\infty p_n[n+1]\int_{\omega_{n}}^\infty\frac{\sin^2(\Omega_{n}t/2)}{\Omega_{n}\sqrt{\Omega_{n}^2-\omega_{n}^2}}\text{d}\Omega_{n}~~~~
\end{eqnarray}
where $p_n(\omega_0)=p_n=\frac{\bar{n}^n}{n!}\text{e}^{-\bar{n}}$ represents the probability of occupation of $n$ coherent photons for the given average number of photons $\bar{n}$, $\Omega_{n}=\pm\sqrt{(\omega-\omega_0)^2+4g_{\omega_0}'^2(n+1)}$ is the renormalization of the generalized $n$-photon Rabi frequency, and $\omega_{n}=2g_{\omega_0}'\sqrt{n+1}$ is the renormalized $n$-photon Rabi frequency \cite{Islam}.

Let us now consider the system be confined to a lossy resonant optical cavity of the resonance frequency $\omega_0$ and the mode quality factor $Q$. The two-level system is interacting with the coherent photons which are injected through a hole on the cavity axis. Let the average number of coherent photons injected on the two-level system be $\bar{n}$. The probability that a photon, which is emitted from the two-level system, escapes from the cavity through the curved surface of the cylindrical-shaped open cavity (of circular mirrors of radius $r$ each and separation $h$) is $p_0=\frac{2\pi rh}{2\pi rh+2\pi r^2}=\frac{1}{1+\frac{r}{h}}$ \cite{Islam}. This probability results in the net quality factor of the cavity as $Q'=\frac{1}{\frac{1}{Q}+\frac{p_0A(0)}{\omega_0}}$ \cite{Islam} where $A(0)$ is the frequency broadening due to the natural decay in the space inside the cavity. Thus we get the net transition probability of two-level system from the excited state ($\ket{\psi_2}$) to the ground state ($\ket{\psi_1}$) at time $t$, by generalizing Eqn. (\ref{eqn:1d}), as \cite{Islam}
\begin{eqnarray}\label{eqn:11}
P_{2\rightarrow1}(t)&=&A(0)\sum_{n=0}^\infty \frac{4}{\pi}\frac{\bar{n}^n}{n!}\text{e}^{-\bar{n}}[n+1]\times\nonumber\\&&\int_{\omega_{n}}^\infty\frac{(\omega_0/Q')^2}{4[\Omega_{n}^2-\omega_{n}^2]+({\omega_0}/{Q'})^2}\frac{\sin^2(\Omega_{n}t/2)}{\Omega_{n}\sqrt{\Omega_{n}^2-\omega_{n}^2}}\text{d}\Omega_{n}\nonumber\\
\end{eqnarray}
where the Lorentzian broadening term $\frac{2}{\pi}\frac{(\omega_0/Q')^2}{4(\Omega_{n}^2-\omega_{n}^2)+(\frac{\omega_0}{Q'})^2}$ takes care of the ``Ohmic" loss to the walls of the cavity, the leakage from the cavity, and the loss due to the spontaneous emission through the open surface of the cavity. Eqn.~(\ref{eqn:11}) is the same as Eqn.~(\ref{eqn:1}) and is a special case of Eqn.~(8) of Ref. \cite{Islam}. Here we have re-derived Eqn.~(\ref{eqn:1}) in a short-cut method.

\end{document}